\documentclass[a4paper,fleqn,usenatbib]{mnras}

\usepackage{newtxtext,newtxmath}

\usepackage[T1]{fontenc}
\usepackage{ae,aecompl}

\usepackage{graphicx}	
\usepackage{amsmath}	
\usepackage{amssymb}	

\usepackage[flushleft]{threeparttable}
\usepackage[pdftex,usenames,dvipsnames]{xcolor}
\usepackage{natbib}
\usepackage{aas_macros}
\usepackage{url}
\usepackage[none]{hyphenat}
\usepackage{times}
\usepackage{array}
\title[Radio jets in dwarf galaxies]{Radio jets from AGN in dwarf galaxies in the COSMOS survey: mechanical feedback out to redshift $\sim$3.4}

\author[Mezcua et al.]{
M. Mezcua$^{1,2}$\thanks{E-mail: marmezcua.astro@gmail.com}, H. Suh$^{3}$\thanks{Subaru fellow}, F. Civano$^{4}$
\\
$^{1}$ Institute of Space Sciences (ICE, CSIC), Campus UAB, Carrer de Magrans, 08193 Barcelona, Spain\\
$^{2}$ Institut d'Estudis Espacials de Catalunya (IEEC), Carrer Gran Capit\`{a}, 08034 Barcelona, Spain\\
${3}$ Subaru Telescope, National Astronomical Observatory of Japan (NAOJ), National Institutes of Natural Sciences (NINS), \\650 North A'ohoku place, Hilo, HI 96720, USA\\
${4}$ Harvard-Smithsonian Center for Astrophysics, Cambridge, MA 02138, USA \\
}

\date{Accepted XXX. Received YYY; in original form ZZZ}

\pubyear{2018}

\begin{document}
\label{firstpage}
\pagerange{\pageref{firstpage}--\pageref{lastpage}}
\maketitle

\begin{abstract}
Dwarf galaxies are thought to host the remnants of the early Universe seed black holes (BHs) and to be dominated by supernova feedback. However, recent studies suggest that BH feedback could also strongly impact their growth. We report the discovery of 35 dwarf galaxies hosting radio AGN out to redshift $\sim$3.4, which constitutes the highest-redshift sample of AGN in dwarf galaxies. The galaxies are drawn from the VLA-COSMOS 3 GHz Large Project and all are star-forming. After removing the contribution from star formation to the radio emission, we find a range of AGN radio luminosities of $L^\mathrm{AGN}_\mathrm{1.4 GHz} \sim 10^{37}$-$10^{40}$ erg s$^{-1}$. The bolometric luminosities derived from the fit of their spectral energy distribution are $\gtrsim 10^{42}$ erg s$^{-1}$, in agreement with the presence of AGN in these dwarf galaxies. The 3 GHz radio emission of most of the sources is compact and the jet powers range from $Q_\mathrm{jet} \sim 10^{42}$ to 10$^{44}$ erg s$^{-1}$. These values, as well as the finding of jet efficiencies $\geq 10$ \% in more than 50\% of the sample, indicate that dwarf galaxies can host radio jets as powerful as those of massive radio galaxies whose jet mechanical feedback can strongly affect the formation of stars in the host galaxy. 
We conclude that AGN feedback can also have a very strong impact on dwarf galaxies, either triggering or hampering star formation and possibly the material available for BH growth. This implies that those low-mass AGN hosted in dwarf galaxies might not be the untouched relics of the early seed BHs, which has important implications for seed BH formation models.

\end{abstract}




\section{Introduction}

The last decade has seen a blossom of detections of active galactic nuclei (AGN) in dwarf galaxies, usually defined as having a stellar mass $M_\mathrm{*} \leq 3 \times 10^{9}$ M$_{\odot}$ (comparable or smaller than that of the Large Magellanic Cloud). This suggests that black holes (BHs) of either intermediate mass ($100 < M_\mathrm{BH} \lesssim 10^{6}$ M$_{\odot}$; IMBHs) or supermassive ($M_\mathrm{BH} > 10^{6}$ M$_{\odot}$) are present in all galaxies and not only in the most massive ones (e.g. \citealt{2013ARA&A..51..511K}; \citealt{2016ASSL..418..263G}). 

Given that obtaining a dynamical mass measurement of an IMBH in a dwarf galaxy is currently limited to slightly beyond the Local Group (e.g. \citealt{2001AJ....122.2469G}; \citealt{2017ApJ...836..237N,2018ApJ...858..118N,2019ApJ...872..104N}), most IMBH searches in dwarf galaxies have focused on finding accretion signatures from low-mass AGN ($M_\mathrm{BH} \lesssim 10^{6}$ M$_{\odot}$) either in the form of high-ionisation optical and infrared emission lines (e.g. \citealt{2008ApJ...677..926S}; \citealt{2015MNRAS.454.3722S}; \citealt{2017A&A...602A..28M}), broad optical emission lines and reverberation mapping techniques (e.g. \citealt{2004ApJ...610..722G,2007ApJ...670...92G}; \citealt{2013ApJ...775..116R}; \citealt{2018ApJ...863....1C}; \citealt{2019arXiv190500145W}) or the detection of X-ray emission brighter than that coming from star formation (e.g. \citealt{2013ApJ...773..150S}; \citealt{2015ApJ...805...12L}; \citealt{2016ApJ...817...20M,2018MNRAS.478.2576M}; see review by \citealt{2017IJMPD..2630021M}). In the case of the optical searches, BH masses are estimated from the width of broad emission lines under the assumption that the ionised gas around the BH is virialised. Most of these optical samples are however incomplete, limited to $z <$ 0.3, or skewed toward high accretion rates (i.e. high Eddington ratios) and type 1 AGN (those with broad emission lines). Moreover, unless the broad line emission is persistent over a time range of $\gtrsim$10 years, its origin might be from transient stellar processes such as supernovae (\citealt{2016ApJ...829...57B}).

The detection of X-ray emission can confirm the presence of accreting BHs in those sources with broad line emission as well as constrain the radiative properties of the population of low-mass AGN (e.g. \citealt{2015ApJ...809L..14B,2017ApJ...836...20B}; \citealt{2018ApJ...863....1C}). To circumvent the biases of optical samples a few research groups have made use of deep X-ray surveys (e.g. \citealt{2015ApJ...799...98M}; \citealt{2016ApJ...831..203P}; \citealt{2018MNRAS.478.2576M}) and estimated the BH mass using either BH-galaxy scaling relations (e.g. $M_\mathrm{BH}-M_\mathrm{*}$; \citealt{2015ApJ...813...82R}) or the fundamental plane of BH accretion (e.g. \citealt{2003MNRAS.345.1057M}; \citealt{2004A&A...414..895F}; \citealt{2006A&A...456..439K}; \citealt{2009ApJ...706..404G,2014ApJ...788L..22G}; \citealt{2012MNRAS.419..267P}; \citealt{2018A&A...616A.152S}; \citealt{2019ApJ...871...80G}). The most complete X-ray studies of low-mass AGN in dwarf galaxies are those we made in \cite{2016ApJ...817...20M}, in which by stacking the X-ray images of $\sim$50,000 dwarf galaxies we found that a population of IMBHs exists in dwarf galaxies at least out to $z$ = 1.5, and \cite{2018MNRAS.478.2576M}, where we reported the highest-redshift discovery of an AGN in a dwarf galaxy at $z$ = 2.4. The large sample of AGN dwarf galaxies found in \cite{2018MNRAS.478.2576M} also allowed us to derive their AGN fraction down to a stellar mass range ($10^{7}-3 \times 10^{9}$ M$_{\odot}$) barely studied before. IMBHs in dwarf galaxies are thought to be the ungrown relics of the high-$z$ seed BHs invoked to explain the existence of quasars at $z\sim$7 (e.g. \citealt{2011Natur.474..616M}; \citealt{2015Natur.518..512W}; \citealt{2018Natur.553..473B}) and of ultramassive BHs (heavier than $10^{10}$ M$_{\odot}$) in the local Universe (e.g. \citealt{2011Natur.480..215M}; \citealt{2018MNRAS.474.1342M}). The fraction of dwarf galaxies hosting IMBHs (for which the AGN fraction is a proxy) can thus provide valuable clues on how the early seed BHs formed (e.g. \citealt{2018MNRAS.tmp.2337R}): simulations predict that a large fraction (90\%) of today's dwarf galaxies should host 'light' ($10^{2}-10^{3}$ M$_{\odot}$) IMBHs if seed BHs formed from the first generation of Population III stars, which are thought to have been very abundant in the early Universe, while a lower occupation fraction (50\%) of 'heavy' ($10^{4}-10^{5}$ M$_{\odot}$) BHs is expected if seed BHs formed from direct gas collapse, a process thought to have been not so common (\citealt{2010MNRAS.408.1139V}; \citealt{2012NatCo...3E1304G}; see review by \citealt{2018arXiv181012310W}). The finding in \cite{2018MNRAS.478.2576M} that the AGN fraction decreases with decreasing stellar mass favors the direct collapse scenario for the formation of seed BHs (e.g. \citealt{2008MNRAS.383.1079V,2016MNRAS.460.2979V}; \citealt{2011ApJ...742...13B,2019MNRAS.482.2913B}; \citealt{2017MNRAS.468.3935H}) while its putative decrease with redshift, a behaviour that differs from that observed in massive galaxies, suggests that BH growth is quenched in dwarf galaxies (e.g. \citealt{2018MNRAS.474.1225A}).

The premise behind dwarf galaxies hosting the leftover of the early seed BHs is that dwarf galaxies have presumably not significantly grown through merger and accretion and are very likely to resemble the primordial galaxies of the infant Universe (e.g. \citealt{2008MNRAS.383.1079V}; \citealt{2010MNRAS.408.1139V}; \citealt{2017MNRAS.468.3935H}). Theoretical models often rely on dwarf galaxies being regulated by stellar outflows (i.e. supernova feedback), which evacuate gas from the nucleus and hamper BH growth (e.g. \citealt{2015MNRAS.452.1502D}; \citealt{2017MNRAS.472L.109A}; \citealt{2017MNRAS.468.3935H}). Because of this supernova feedback, in dwarf galaxies neither the galaxy nor the BH are thought to have grown much over cosmic time, providing a promising laboratory where the mass of the BH is expected not to differ much from its initial mass (e.g. \citealt{2008MNRAS.383.1079V,2010MNRAS.408.1139V}). This would explain why low-mass galaxies tend to deviate from the BH-galaxy scaling relations of massive galaxies (e.g. \citealt{2017IJMPD..2630021M}; \citealt{2018ApJ...855L..20M}; \citealt{2018ApJ...863....1C}; \citealt{2019MNRAS.485.1278S}), whose growth is thought to be entwined with that of their central BHs via galaxy mergers (e.g. \citealt{2013ARA&A..51..511K}). Alternatively, the change of slope at the low-mass end of the scaling relations could be also attributed to a bimodality in the accretion efficiency of seed BHs (\citealt{2018ApJ...864L...6P}).

Some recent simulations (\citealt{2016MNRAS.463.2986S}; \citealt{2018MNRAS.473.5698D}; \citealt{2018arXiv180704768B}; \citealt{2019MNRAS.484.2047K}; \citealt{2019MNRAS.486.3892R}; \citealt{2019MNRAS.483.1957Z}) and observational studies (\citealt{2018ApJ...861...50B}; \citealt{2018MNRAS.476..979P}; \citealt{2019arXiv190201401D}) indicate that BH feedback can also have a big impact on dwarf galaxies. Stellar tidal disruption events could for instance fuel BH accretion in dwarf galaxies and their outflows have observable effects on the host galaxies (\citealt{2019MNRAS.483.1957Z}), while outflows from AGN winds (\citealt{2018MNRAS.473.5698D}) or generated by bipolar jets (\citealt{2019MNRAS.486.3892R}) could suppress BH accretion and be a more significant source of feedback than supernovae. AGN feedback can not only quench star formation and BH accretion but also trigger the formation of stars (e.g. \citealt{2009ApJ...700..262S}; \citealt{2012MNRAS.425..438G}; \citealt{2012MNRAS.427.2401K,2014MNRAS.442.1181K}; \citealt{2013MNRAS.433.3297D}; \citealt{2013ApJ...772..112S}; \citealt{2016A&A...593A.118Q}; \citealt{2017Natur.544..202M}), so that in dwarf galaxies hosting AGN the growth of the BH might not be stunted but enhanced by AGN feedback (\citealt{2019NatAs...3....6M}). This has momentous implications for understanding how seed BHs form, since if IMBHs in local dwarf galaxies have significantly grown through AGN feedback then they should not be treated as the leftover of the early Universe seed BHs (\citealt{2019NatAs...3....6M}).

Observational examples of AGN feedback in action in a dwarf galaxy are however scant (e.g. \citealt{2017ApJ...845...50N}; \citealt{2018ApJ...861...50B}; \citealt{2018MNRAS.476..979P}; \citealt{2019arXiv190201401D}; \citealt{2019arXiv190509287M}) and only in about ten IMBHs/low-mass AGN have (radio) jets been detected (\citealt{2006ApJ...636...56G}; \citealt{2006ApJ...646L..95W}; \citealt{2008ApJ...686..838W}; \citealt{2011AN....332..379M}; \citealt{2012ApJ...753..103N}; \citealt{2012ApJ...750L..24R}; \citealt{2012Sci...337..554W}; \citealt{2013MNRAS.436.1546M,2013MNRAS.436.3128M}; \citealt{2014ApJ...787L..30R}; \citealt{2015MNRAS.448.1893M,2018MNRAS.478.2576M,2018MNRAS.480L..74M}). With the advent of the next generation Very Large Array, jet radio emission will be detectable from 10$^{4}$ M$_{\odot}$ BHs at the distance of the Virgo Cluster or from 10$^{6}$ M$_{\odot}$ BHs out to 1 Gpc (\citealt{2018ASPC..517..719P}). These distances can be boosted by deep radio surveys, which will allow us to detect AGN radio emission in dwarf galaxies at very high redshifts. Based on the VLA-COSMOS 3 GHz Large Project (\citealt{2017A&A...602A...1S}), in this paper we report the largest sample of radio AGN dwarf galaxies to date (35 sources). With detection up to $z$ = 3.4, this sample constitutes the highest-redshift discovery of AGN in dwarf galaxies, beating the previous record-holder at $z$ = 2.4 (\citealt{2018MNRAS.478.2576M}). The high jet efficiencies found for most of these radio AGN dwarf galaxies indicate that their jets are, as in massive radio galaxies, able to impart mechanical feedback, bolstering the recent findings that AGN feedback might play a significant role in dwarf galaxies. 

The sample selection and analysis are described in Section~\ref{sample}, while the results obtained are reported and discussed in Section~\ref{discussion}. Final conclusions and open issues are provided in Sect.~\ref{conclusions}. Throughout the paper we adopt a $\Lambda$CDM cosmology with parameters $H_{0}=70$ km s$^{-1}$ Mpc$^{-1}$, $\Omega_{\Lambda}=0.73$ and $\Omega_{m}=0.27$.

\section{Sample and analysis}
\label{sample}
To investigate the presence of radio jets in dwarf galaxies we use the radio source catalog of the VLA-COSMOS 3 GHz Large Project (\citealt{2017A&A...602A...1S}), which is a 2.6 deg$^{2}$ radio continuum survey performed with the Karl G. Jansky Very Large Array (VLA) that encloses the full Cosmic Evolution Survey (COSMOS; \citealt{2007ApJS..172....1S}) field. The COSMOS field has been covered by nearly all major observational facilities (\textit{Hubble}, \textit{Spitzer}, Subaru, Canada-France-Hawaii Telescope, Magellan, VLT, \textit{Herschel}, \textit{GALEX}, \textit{Chandra}, \textit{XMM-Newton}, \textit{NuSTAR}), including the VLA at 1.4 GHz (rms$\sim$10-15 $\mu$Jy beam$^{-1}$, resolution of 1.5 arcsec; \citealt{2007ApJS..172...46S,2010ApJS..188..384S}), constituting one of the largest surveys with a complete multiwavelength dataset. The VLA-COSMOS 3 GHz Large Project reaches a mean rms of 2.3 $\mu$Jy beam$^{-1}$, an angular resolution of 0.75 arcsec, and contains 10830 sources detected down to 5$\sigma$ (\citealt{2017A&A...602A...1S}). Of these, 93\% have counterparts when crossmatching with the COSMOS ancillary multiwavelength (from optical to sub-millimeter) data (\citealt{2017A&A...602A...2S}; \citealt{2017A&A...602A...3D}). A multiwavelength spectral energy distribution (SED) fitting of the radio sources was performed using AGN and galaxy templates in order to derive integrated galaxy properties, such as stellar mass and star formation rate (SFR), for each individual source (\citealt{2017A&A...602A...3D}). The sources were further classified as either star-forming galaxies, high to moderate radiative luminosity AGN (HLAGN), or moderate to low radiative luminosity AGN (MLAGN) based on a combination of SED, X-ray, radio, and mid-infrared (mid-IR) diagnostics (\citealt{2017A&A...602A...2S}; \citealt{2017A&A...602A...3D}). HLAGN include SED-selected AGN, X-ray AGN (selected as having $L_\mathrm{0.5-8 keV} \geq 10^{42}$ erg s$^{-1}$), and mid-IR AGN (classified as AGN in the mid-IR colour-colour diagram of \citealt{2012ApJ...748..142D}); while MLAGN include radio AGN whose 1.4 GHz radio emission shows a $>3\sigma$ excess relative to the radio emission expected from the SFR of the host galaxy. 

The sample of dwarf galaxies analysed in this paper is drawn from the VLA-COSMOS 3 GHz Large Project radio source catalog as having a signal-to-noise ratio $>5$, stellar masses $7 <$ log ($M_\mathrm{*}$/M$_{\odot}$) $\leq$ 9.5, and classified as either HLAGN or MLAGN. This results in a parent sample of 80 radio-emitting dwarf galaxies hosting AGN. For those sources with a false match probability higher than 20\% we visually inspect the optical and radio image positions and exclude from the sample four sources which seem to be a spurious crossmatch. We also inspect the probability distribution function (PDF) of the stellar masses and remove from the sample two additional sources whose PDFs extend beyond $10^{10}$ M$_{\odot}$. In order to have a sample of dwarf galaxies as bona-fide as possible, for the remaining 74 sources we re-compute the stellar masses and SFRs considering the uncertainties of the photometric redshift from \cite{2016ApJS..224...24L} and \cite{2017A&A...602A...3D}. We use the SED-fitting technique of \cite{2017ApJ...841..102S,2019ApJ...872..168S} to decompose the SED into a dust emission component from a nuclear AGN torus, a host galaxy emission component from stellar populations, and a far-IR emission starburst component. We derive the PDF for the host stellar mass considering any possible combination of SED parameters, which includes the age since the onset of star formation, the e-folding time $\tau$ for exponential star formation history models, and the dust reddening. A full detailed description of the SED fitting method is presented in \cite{2017ApJ...841..102S,2019ApJ...872..168S}. We also compute the monochromatic AGN luminosity at rest-frame 6$\mu$m from the best-fitting AGN torus template and derive the bolometric luminosity using the relation between $L_\mathrm{6\mu m}$ and $L_\mathrm{bol}$ of \cite{2019ApJ...872..168S}. Based on this SED fitting, only 43 of the 74 sources have $7 <$ log ($M_\mathrm{*}$/M$_{\odot}$) $\leq$ 9.5 and can thus be reliably classified as dwarf galaxies. We thus consider this sample of 43 radio-emitting dwarf galaxies for further analysis. Their redshifts range from $z$ = 0.13 to 3.4, seven of them being spectroscopic and the rest photometric (see Table~\ref{radioproperties}), and their 3 GHz radio luminosities from $L_\mathrm{3 GHz}$ = $3.2 \times 10^{21}$ W Hz$^{-1}$ to $1.5 \times 10^{24}$ W Hz$^{-1}$ (see Table~\ref{radioproperties}).

\subsection{Radio spectral index}
\label{spectralindex}
Eight of the 43 dwarf galaxies with radio emission at 3 GHz have also a VLA radio counterpart at 1.4 GHz. For these eight galaxies we are able to compute the radio spectral index $\alpha$ as $S_{\nu} \propto \nu^{-\alpha}$, where $S_{\nu}$ is the radio flux at the frequencies $\nu$= 3 and 1.4 GHz, finding that it ranges from $\alpha$ = 0.3 to 2.8. For the remaining sources a typical spectral index of 0.7 is assumed, consistent with the average value found for the full population of 3 GHz radio sources in the VLA-COSMOS 3 GHz Large Project (\citealt{2017A&A...602A...1S}). The 1.4 GHz luminosities derived from the 1.4 GHz radio flux or from the one at 3 GHz assuming $\alpha = 0.7$ and applying the corresponding K-correction factor range from $3.2 \times 10^{21}$ W Hz$^{-1}$ to $2.6 \times 10^{24}$ W Hz$^{-1}$.

Of the eight dwarf galaxies with 1.4 GHz radio counterparts, four are found to be compact. The spectral index of these four compact sources is steep (>0.7), so they are compact, steep-spectrum sources such as the low-mass AGN GH10 (\citealt{2008ApJ...686..838W}). GH10 is found by \cite{2008ApJ...686..838W} to resemble the Seyfert galaxies of the Palomar survey (\citealt{2001ApJS..133...77H}), which show compact radio cores sometimes slightly resolved into an elongated, jet-like outflow component and a wide range of radio luminosities ($10^{18}-10^{25}$ W Hz$^{-1}$) and spectral indices (+0.5 to -1; \citealt{2001ApJS..133...77H}; \citealt{2001ApJ...558..561U}). 
The 43 dwarf galaxies with radio emission also show a wide range (though slightly narrower than the Palomar Seyferts) of radio luminosities and 35 of them are compact at 3 GHz. Their radio properties, specially of those with 1.4 GHz radio counterparts, are therefore very similar to those of the steep-spectrum Palomar Seyferts, suggesting that their radio emission could be outflow-driven as in GH10.

\subsection{Contribution from star formation}
\label{starformation}
The radio luminosity of the 43 dwarf galaxies is several orders of magnitude higher than those of X-ray binaries and of even the brightest supernova remnants (SNRs) such as SNR N4449-1, with a 1.6 GHz radio luminosity of 1.09 $\times$ 10$^{19}$ W Hz$^{-1}$ (higher than that of the Galactic SNRs Cas A or Crab and comparable to the peak luminosity of SNe such as SN 1970G; \citealt{2013MNRAS.436.2454M}). This makes the 43 dwarf galaxies strong AGN candidates. 

According to the near-ultraviolet (NUV) to r-band colour-colour diagram (e.g. \citealt{2010ApJ...709..644I}), 40 out of the 43 radio-emitting dwarf galaxies have $M_\mathrm{NUV}-M_\mathrm{r} < 3.5$ and are classified as star-forming. The star-forming nature of the dwarf galaxy sample is further supported by their SFR, which is log ($SFR$/M$_\mathrm{\odot}$ yr$^{-1}$) $> -2$ in the 43 dwarf galaxies. Therefore, the detected radio emission is most likely a mixed contribution of emission from star formation and nuclear AGN emission. This is reinforced by the finding that only 7 out of the 43 dwarf galaxies have $L_\mathrm{1.4 GHz} > 5 \times 10^{23}$ W Hz$^{-1}$ and are thus clear radio AGN (e.g. \citealt{2014ApJ...784..137K}), as such high radio luminosities cannot be explained by star formation alone. The contribution from star formation to the radio emission must be thus removed from the detected radio emission in order to probe the presence of AGN jet emission in the dwarf galaxies. 

\begin{figure}
\includegraphics[width=0.49\textwidth]{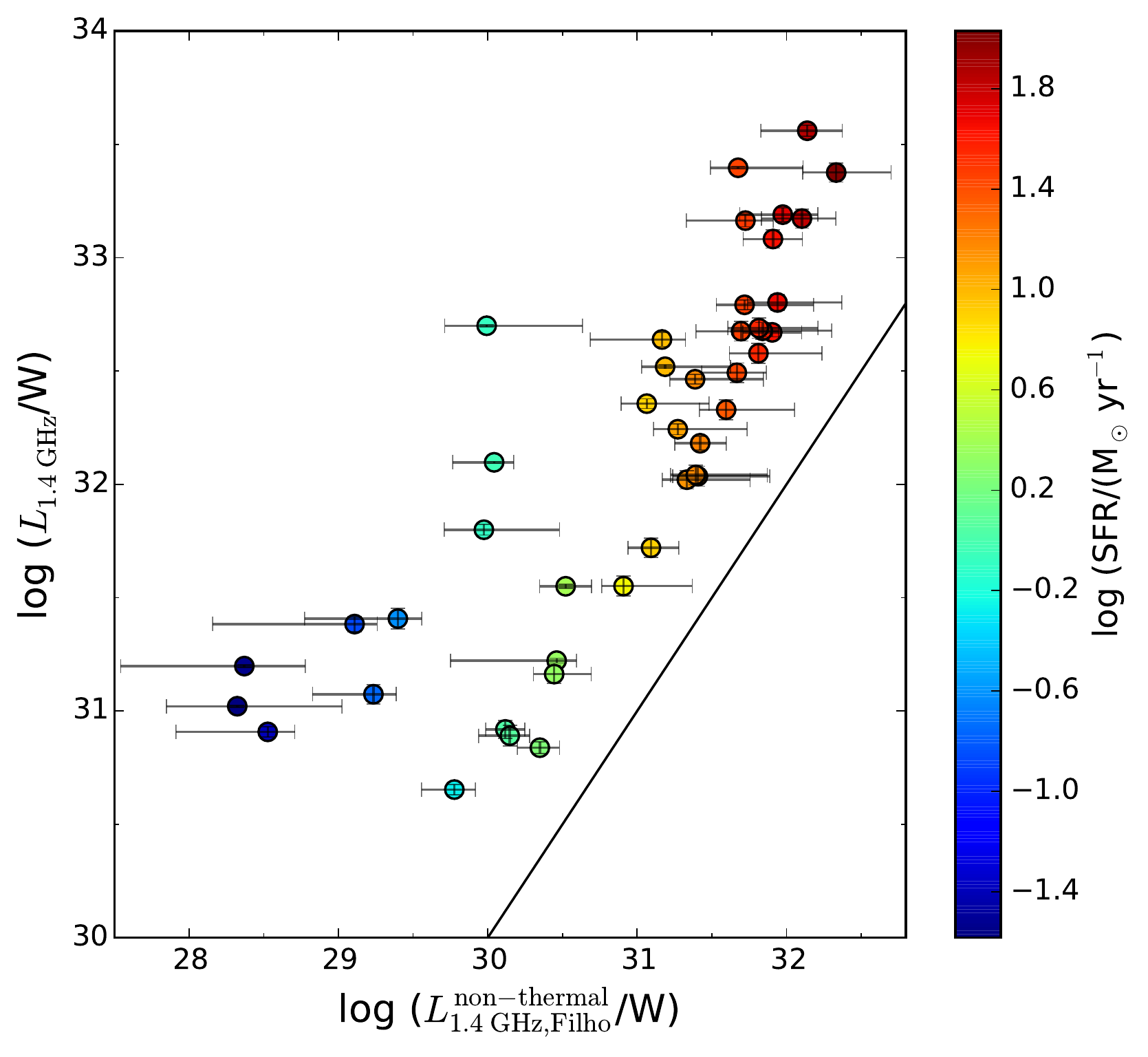}
\includegraphics[width=0.49\textwidth]{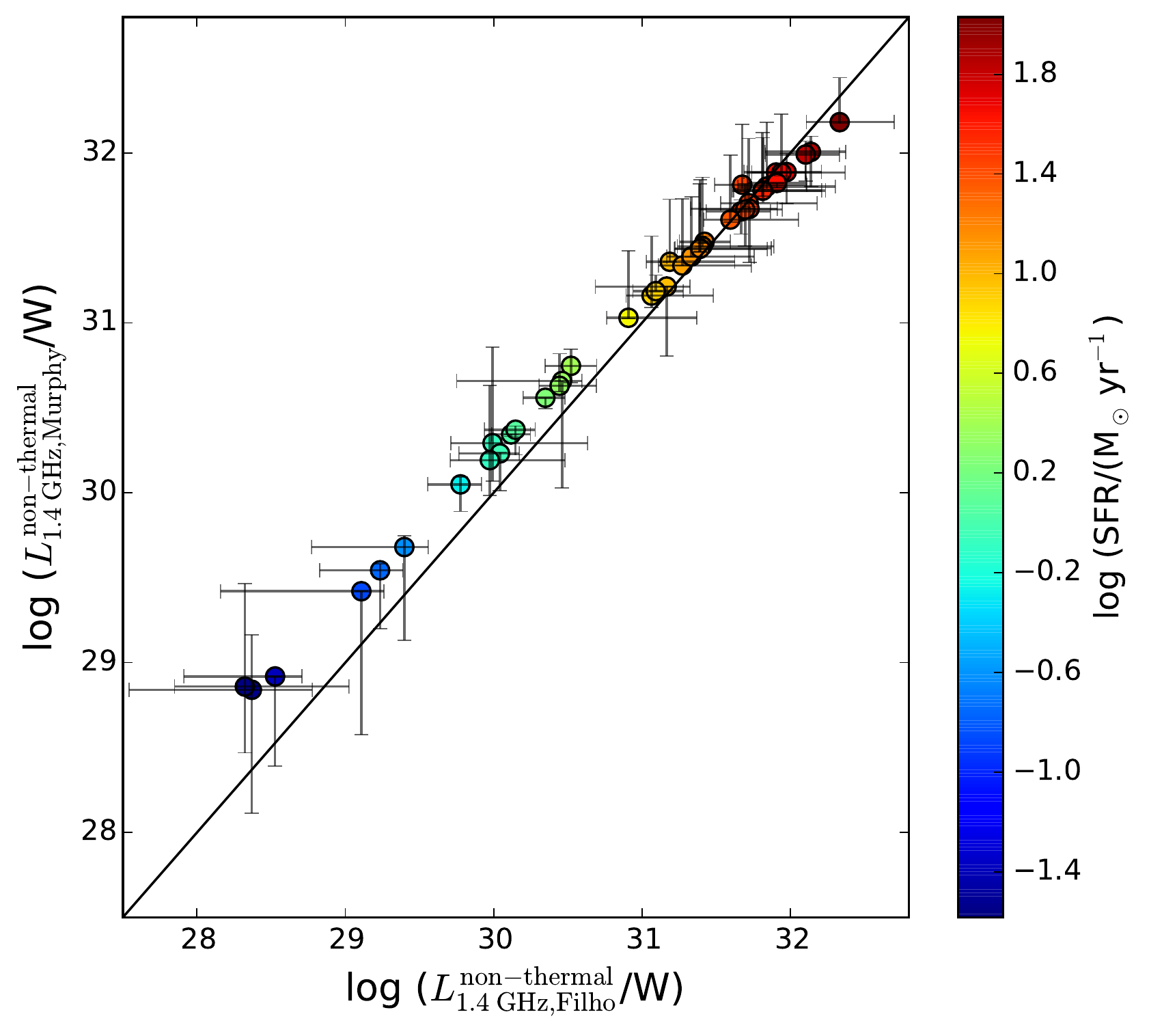}
\protect\caption[figure]{\textbf{Top}. Radio luminosity at 1.4 GHz for the parent sample of 43 dwarf galaxies versus non-thermal 1.4 GHz radio luminosity expected from star formation derived using the correlation for dwarf galaxies of \cite{2019MNRAS.484..543F}. \textbf{Bottom}. Non-thermal 1.4 GHz radio luminosity expected from star formation derived using the global correlation of \cite{2011ApJ...737...67M} versus that obtained using the correlation for dwarf galaxies of \cite{2019MNRAS.484..543F}. In both plots the solid line denotes a one-to-one correlation.}
\label{SF}
\end{figure}

To estimate the radio emission from star formation processes we use the correlation between SFR and non-thermal 1.4 GHz luminosity of \cite{2019MNRAS.484..543F}:
\begin{equation}
\label{filho}
\begin{split}
\mathrm{log (}L_\mathrm{1.4 GHz, Filho}^\mathrm{non-thermal}\mathrm{/W)} = \mathrm{log (} SFR\ \mathrm{/M}_{\odot}\ \mathrm{yr}^{-1}\mathrm{)} \times (1.11 \pm 0.09) + \\(30.08 \pm 0.13)
\end{split}
\end{equation}
derived for a sample of dwarf galaxies, including extremely metal-poor systems, with radio continuum, IR and CO emission data available and stellar masses below $10^{10}$ M$_{\odot}$. The 1$\sigma$ errors on the SFR are also taken into account in the computation of log $L_\mathrm{1.4 GHz, Filho}^\mathrm{non-thermal}$. We find that the contribution of star formation to the 1.4 GHz radio luminosity is lower than 32\% and that the 1.4 GHz radio luminosity is more than 2$\sigma$ larger than that expected from star formation for all sources (see Fig.~\ref{SF}). We note that the radio luminosity no longer traces SFR for radio luminosities below 10$^{27}$ W, or SFR $\leq$ 0.01 M$_{\odot}$ yr$^{-1}$ (i.e. the correlation steepens at low luminosities), but this does not affect our sample of dwarf galaxies as the 43 sources have SFR $\geq$ 0.03 M$_{\odot}$ yr$^{-1}$. To reinforce this we compare the $L_\mathrm{1.4 GHz,Filho}^\mathrm{non-thermal}$ derived using eq.~\ref{filho} with that obtained using the more general $SFR-L_\mathrm{1.4 GHz}^\mathrm{non-thermal}$ correlation of \cite{2011ApJ...737...67M}:
\begin{equation}
L_\mathrm{1.4 GHz, Murphy}^\mathrm{non-thermal}  \mathrm{(erg\ s}^{-1} \mathrm{Hz}^{-1}\mathrm{)} = 1.57\ SFR\ \mathrm{(M}_{\odot}\ \mathrm{yr}^{-1}\mathrm{)} 
\end{equation}
which is based on a radio far-IR correlation established for globally integrated properties of galaxies.
For the SFR $\geq$ 0.03 M$_{\odot}$ yr$^{-1}$ proben here, the 1.4 GHz luminosities expected from star formation are fully consistent with those derived from the correlation of \cite{2019MNRAS.484..543F} (see Fig.~\ref{SF}, where a slight steepening at low luminosities is started to be seen). 

In addition to the non-thermal emission from star formation, free-free emission from thermal processes could also contribute to the radio emission. The thermal fraction in spiral galaxies is typically of 10\%, while in dwarf galaxies it is generally no more than 30\% (e.g. \citealt{1992ARA&A..30..575C}; \citealt{2012MNRAS.423L.127R}; \citealt{2019MNRAS.484..543F}). To estimate the contribution of thermal emission to the 1.4 GHz radio luminosity ($L_\mathrm{1.4 GHz}^\mathrm{thermal}$) we use the nearly one-to-one ratio between thermal and non-thermal emission of  \cite{2019MNRAS.484..543F}: log $L_\mathrm{1.4 GHz, Filho}^\mathrm{non-thermal} \propto$ 0.92 $\times$ log $L_\mathrm{1.4 GHz}^\mathrm{thermal}$. This results in a thermal fraction $\lesssim$ 32\%. Considering both the thermal and non-thermal contribution, we find that the 1.4 GHz radio luminosity is still $\geq2\sigma$ larger than that expected from star formation for all sources. 

Although all the sources are classified as either MLAGN or HLAGN by \cite{2017A&A...602A...2S} and \cite{2017A&A...602A...3D} and the 1.4 GHz radio luminosity is $\geq2\sigma$ larger than that expected from star formation for the 43 dwarf galaxies in the parent sample, to investigate the presence of jet radio emission we select only those sources whose 1.4 GHz radio luminosity is more than 3$\sigma$ larger than that coming from star formation. This is the case for 35 of the 43 dwarf galaxies. For these 35 radio AGN dwarf galaxies we then compute the AGN radio luminosity by subtracting the thermal and non-thermal luminosities derived above from the 1.4 GHz radio luminosity. 

\section{Results and discussion}
\label{discussion}
\subsection{Radio AGN host galaxies}
\begin{figure*}
\includegraphics[width=\textwidth]{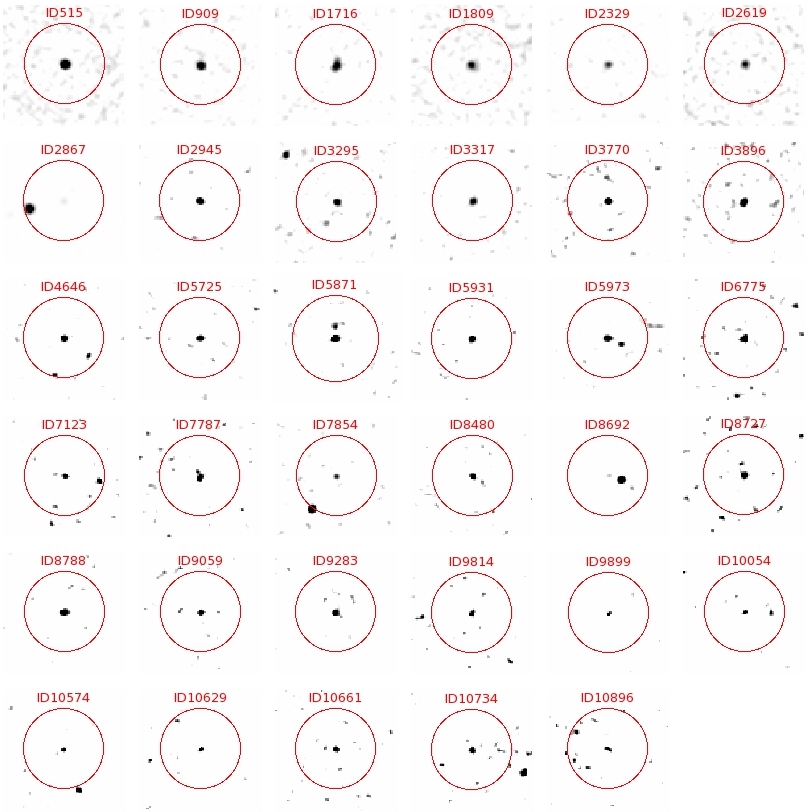}
\protect\caption[figure]{VLA cutouts of the 35 radio AGN dwarf galaxies. The red circles mark the optical position with a radius of 2 arcsec.}
\label{mosaic}
\end{figure*}

\begin{figure}
\includegraphics[width=0.48\textwidth]{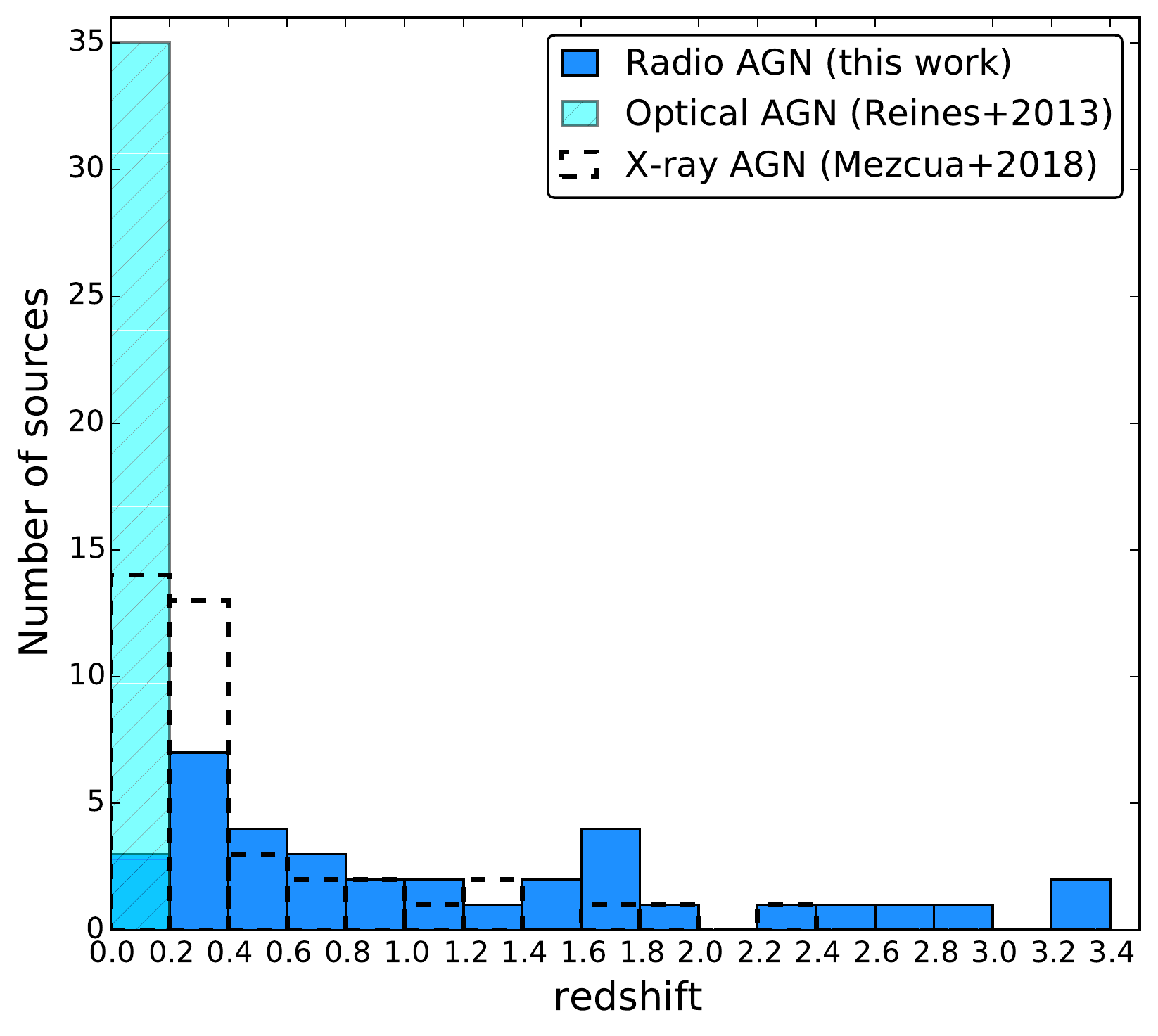}
\includegraphics[width=0.49\textwidth]{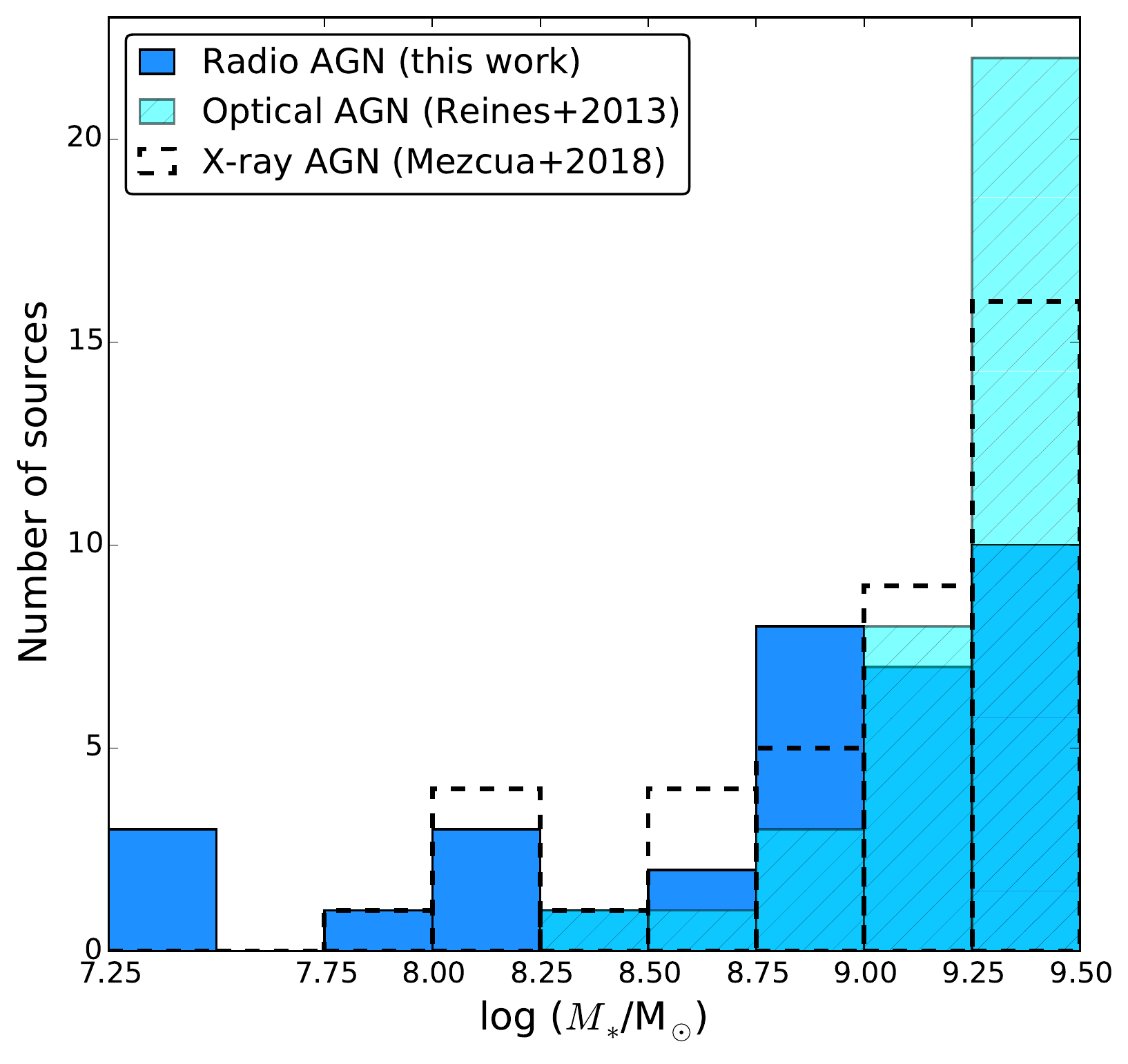}
\protect\caption[figure]{Redshift (top) and stellar mass (bottom) distribution for the sample of 35 radio AGN dwarf galaxies (blue solid bars), the optically-selected AGN sample of \cite{2013ApJ...775..116R} (cyan hashed bars), and the X-ray sample of AGN dwarf galaxies from \cite{2018MNRAS.478.2576M} (dashed bars).}
\label{redshift}
\end{figure}

The final sample of 35 radio AGN dwarf galaxies have 1.4 GHz AGN radio luminosities ranging from $L^\mathrm{AGN}_\mathrm{1.4 GHz}$ = 3.7 $\times$ 10$^{37}$ erg s$^{-1}$ to 3.4 $\times$ 10$^{40}$ erg s$^{-1}$ (see Table~\ref{radioproperties}). Their VLA cutouts are shown in Fig.~\ref{mosaic}. The redshift of these 35 radio AGN dwarf galaxies ranges from $z$ = 0.13 to 3.4 (see Fig.~\ref{redshift}, top). With 4 sources at $z > $2.5, this constitutes the highest-redshift discovery of AGN in dwarf galaxies. Previous optically-selected samples of AGN in dwarf galaxies were limited to the local Universe (e.g. \citealt{2013ApJ...775..116R}; \citealt{2018ApJ...863....1C}), which was circumvented only by X-ray searches such as that from \cite{2016ApJ...831..203P} and \cite{2016ApJ...817...20M,2018MNRAS.478.2576M}. Using the \textit{Chandra} COSMOS Legacy Survey (\citealt{2016ApJ...819...62C}), \cite{2016ApJ...817...20M} found that a population of IMBHs must exist in dwarf galaxies at least out to $z$ = 1.5, which was later confirmed by the finding of a sample of X-ray AGN out to $z$ = 2.39 (\citealt{2018MNRAS.478.2576M}). The new redshift record-holder found here as a radio AGN is COSMOSVLA3J095934.30+014735.9, or ID\_VLA 3896 (see Table~\ref{radioproperties}), at a photometric redshift $z$ = 3.4$^{+0.2}_{-0.3}$ and with $L^\mathrm{AGN}_\mathrm{1.4 GHz}$ = (3.4 $\pm$ 0.4) $\times$ 10$^{40}$ erg s$^{-1}$. 

When going from optical searches to X-ray and radio surveys there is an increase on the highest redshift of the AGN dwarf galaxies detected but also a decrease on the lowest detected stellar mass (see Fig.~\ref{redshift}, bottom). The use of X-ray and radio surveys such as the \textit{Chandra} COSMOS Legacy Survey and VLA-COSMOS 3 GHz Large Project have allowed the detection of AGN in several dwarf galaxies below 10$^{8}$ M$_{\odot}$ (e.g. \citealt{2018MNRAS.478.2576M}). The new record-holders for the lightest dwarf galaxies to host an AGN are ID\_VLA 2867 and 8727 (see Fig.~\ref{redshift}, bottom), found in this paper as radio AGN at $z < 0.4$ with log ($M_\mathrm{*}$/M$_{\odot}$) = 7.4 $\pm$ 0.8 and log ($M_\mathrm{*}$/M$_{\odot}$) = 7.4 $\pm$ 0.7 respectively, and $L^\mathrm{AGN}_\mathrm{1.4 GHz}$ = (1.58 $\pm$ 0.05) $\times$ 10$^{38}$ erg s$^{-1}$ and (6.5 $\pm$ 1.5) $\times$ 10$^{37}$ erg s$^{-1}$, respectively.

\begin{figure}
\includegraphics[width=0.49\textwidth]{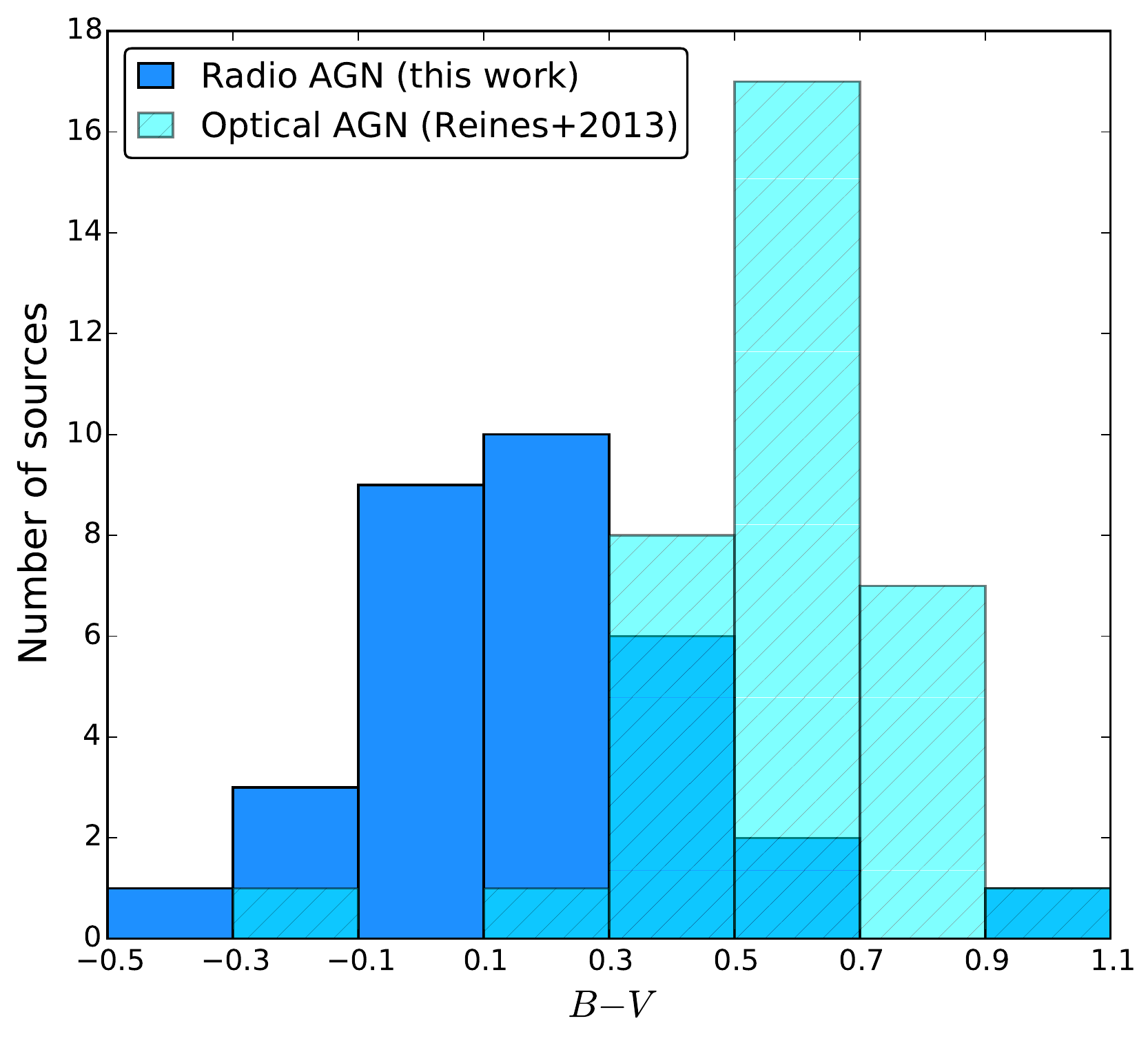}
\protect\caption[figure]{Distribution of the $B-V$ colour for the sample of 35 radio AGN dwarf galaxies (blue solid bars) and the optically-selected AGN sample of \cite{2013ApJ...775..116R} (cyan hashed bars).}
\label{color}
\end{figure}

The host galaxies of the radio AGN have a <$B-V$> colour of 0.18, indicative of blue galaxies and consistent with the star-forming nature (Sect.~\ref{starformation}). Such colours are bluer than those of optically-selected AGN (see Fig.~\ref{color}), which tend to be redder. This is most likely a selection effect of optical spectroscopic diagnostics, which, unlike X-ray and radio searches, are not able to identify AGN in star-forming galaxies (e.g. \citealt{2013ApJ...775..116R}).

From the B-band optical flux and 5 GHz AGN radio flux we also derive the radio loudness $R$ ($R$ = $f_\mathrm{5 GHz}/f_{4400 \dot{A}}$, where $R <$ 10 means radio-quiet; \citealt{1989AJ.....98.1195K}) of the 35 radio AGN dwarf galaxies. The 5 GHz radio flux is derived from the 1.4 GHz AGN one properly scaled using the radio spectral indexes derived in Sect. \ref{spectralindex}. We find a radio-loud fraction of 63\%, much higher than those of previous samples of low-mass AGN with detected radio emission (e.g. 1\%, \citealt{2007ApJ...670...92G}; 4 \%, \citealt{2018ApJS..235...40L}) and than that of classical quasars and AGN (15\%-20\%, e.g. \citealt{2002AJ....124.2364I}; \citealt{2006ApJ...636...56G}; \citealt{2015ApJ...804..118B}).

\begin{table*}
\begin{minipage}{\textwidth}
\centering
\caption{Radio AGN dwarf galaxy properties.}
\label{radioproperties}
\begin{tabular}{lcccccccccc}
\hline
\hline 
  ID$\_$VLA &      RA(J2000)    &     DEC(J2000) &  z            &  log $M_\mathrm{*}$	&  	i	&  log $SFR$  				&  log $L_\mathrm{bol}$ & Int. flux (3 GHz) &	log $L^\mathrm{AGN}_\mathrm{1.4GHz}$  & $Q_\mathrm{jet}$  	\\
  	 	 & 		(deg)  	&  (deg) 		& 		   &  (M$_\mathrm{\odot}$) &  (mag)	&(M$_\mathrm{\odot}$ yr$^{-1}$) &   (erg s$^{-1}$)	  &     	($\mu$Jy)		&  (erg s$^{-1}$)	                  &        	(erg s$^{-1}$)		\\
    (1)	    	&   		(2)     	  &     (3)  	 	&   	(4)        &  	(5)		  	        &	  (6)  &    (7)   					 &  		  (8)  		  & 		(9)  			&			(10) 			&	(11)	 \\    
 \hline
      515  &  150.496858  &  2.473441  &  0.52$^{+0.04}_{0.13}$$^{p}$ & 8.5$^{+0.6}_{0.6}$  &        25.19  &  -0.1$^{+0.2}_{0.6}$  & 43.6$^{+1.0}_{1.0}$ &  131.9$\pm$7.4  &      39.70$\pm$0.01  &  43.6  \\
      909  &  149.479617  &  1.972097  &  0.29$^{+0.01}_{0.01}$$^{p}$ & 9.2$^{+0.2}_{0.3}$  &        22.90  &  -0.0$^{+0.2}_{0.0}$  & 43.8$^{+0.9}_{0.9}$ &  123.1$\pm$7.0  &     39.09$\pm$0.01  &  43.1  \\
     1716  &  149.968248  &  2.675955  &  1.19$^{+0.29}_{0.14}$$^{p}$ &  9.0$^{+0.8}_{0.8}$  &        26.31  &   1.4$^{+0.0}_{0.4}$  & 44.1$^{+1.0}_{1.0}$ &   75.3$\pm$4.4  &     40.38$\pm$0.01  &  44.0  \\
     1809  &  150.557523  &  2.330501  &  0.29$^{+0.01}_{0.01}$$^{p}$ &  8.2$^{+0.6}_{0.6}$  &        24.52  &   0.4$^{+0.1}_{0.1}$  & 43.5$^{+0.9}_{0.9}$ &   53.7$\pm$3.5  &     38.49$\pm$0.02  &  42.7  \\
     2329  &  149.904547  &  2.552293  &  0.76$^{+0.05}_{0.04}$$^{p}$ &  9.5$^{+0.2}_{0.2}$  &        24.46  &   1.0$^{+0.0}_{0.4}$  & 43.7$^{+1.0}_{0.9}$ &   36.6$\pm$3.0  &      39.49$\pm$0.02  &  43.4  \\
     2619  &  150.256994  &  2.211813  &    1.18$^{+0.00}_{0.00}$$^{s}$ &9.1$^{+0.1}_{2.9}$  &        23.98  &   1.6$^{+0.0}_{0.0}$  &  46.1$^{+0.9}_{0.9}$ &   30.0$\pm$2.7  &     39.54$\pm$0.05  &  43.4  \\
     2867  &  149.611112  &  2.620181  &   0.18$^{+0.22}_{0.1}$ $^{p}$ &7.4$^{+0.8}_{0.8}$  &        25.21  &  -1.5$^{+0.7}_{0.3}$  &  42.3$^{+0.9}_{1.0}$  &   36.2$\pm$3.4  &      38.20$\pm$0.01  &  42.5  \\
     2945  &  149.871162  &  1.620135  &  0.95$^{+0.12}_{0.06}$$^{p}$ & 8.8$^{+0.6}_{0.6}$  &        25.42  &   1.2$^{+0.0}_{0.4}$  & 44.2$^{+0.9}_{1.1}$ &   31.3$\pm$3.0  &     39.41$\pm$0.05  &  43.3  \\
     3295  &  150.338442  &  1.744434  &  1.42$^{+0.17}_{0.15}$$^{p}$ &  8.9$^{+0.6}_{0.6}$  &        25.68  &   1.5$^{+0.0}_{0.4}$  & 44.5$^{+0.9}_{0.9}$ &   25.9$\pm$2.6  &     39.73$\pm$0.05  &  43.6  \\
     3317  &  149.760790  &  2.622779  &  0.22$^{+0.01}_{0.01}$$^{p}$ &  8.2$^{+0.2}_{0.1}$  &        22.87  &  -1.4$^{+0.5}_{0.0}$  & 42.8$^{+0.9}_{0.9}$ &   26.1$\pm$2.7  &     37.91$\pm$0.05  &  42.3  \\
     3770  &  149.635991  &  2.403578  &   0.85$^{+0.1}_{0.12}$$^{p}$ & 8.7$^{+0.8}_{0.6}$  &        26.56  &   1.1$^{+0.0}_{0.4}$  & 43.1$^{+1.0}_{0.9}$ &   24.8$\pm$2.7  &     39.17$\pm$0.06  &  43.2  \\
     3896  &  149.892936  &  1.793327  &  3.35$^{+0.17}_{0.26}$$^{p}$ &  9.1$^{+0.5}_{0.5}$  &        26.09  &   1.9$^{+0.2}_{0.1}$  & 45.5$^{+0.9}_{0.9}$ &   22.0$\pm$2.5  &     40.53$\pm$0.05  &  44.1  \\
     4646  &  149.867262  &  2.349574  &  2.26$^{+0.14}_{0.16}$$^{p}$ &  9.4$^{+0.5}_{0.5}$  &        24.93  &   1.5$^{+0.3}_{0.0}$  & 45.0$^{+0.9}_{0.9}$ &   21.1$\pm$2.6  &     40.14$\pm$0.06  &  43.9  \\
     5725  &  150.810928  &  2.689113  &  0.78$^{+1.32}_{0.56}$$^{p}$ & 8.8$^{+1.1}_{1.0}$  &        25.89  &   1.2$^{+0.0}_{0.0}$  &  44.9$^{+0.9}_{0.9}$  &   25.8$\pm$3.6  &      39.06$\pm$0.08  &  43.1  \\
     5871  &  150.779415  &  1.699318  &   1.3$^{+0.07}_{0.08}$$^{p}$ & 9.3$^{+0.7}_{0.6}$  &        26.89  &   1.0$^{+0.4}_{0.0}$  & 45.3$^{+0.9}_{0.9}$ &   22.6$\pm$3.2  &     39.62$\pm$0.06  &  43.5  \\
     5931  &  149.769009  &  2.154255  &  2.49$^{+0.16}_{0.23}$$^{p}$ &  9.3$^{+0.5}_{0.5}$  &        24.88  &   1.7$^{+0.2}_{0.1}$  & 45.1$^{+0.9}_{0.9}$ &   18.0$\pm$2.5  &     40.15$\pm$0.07  &  43.9  \\
     5973  &  150.424442  &  2.453789  &    0.17$^{+0.00}_{0.00}$$^{s}$ &8.8$^{+0.2}_{0.2}$  &        22.19  &  -0.3$^{+0.2}_{0.0}$  &  43.7$^{+0.9}_{0.9}$ &   23.8$\pm$2.5  &     37.57$\pm$0.06  &  42.1  \\
     6775  &  149.891879  &  2.771587  &  0.44$^{+0.01}_{0.01}$$^{p}$ &  8.9$^{+0.3}_{0.3}$  &        24.52  &  -0.9$^{+0.8}_{0.0}$  & 43.4$^{+0.9}_{1.1}$ &   15.5$\pm$2.4  &     38.38$\pm$0.07  &  42.6  \\
     7123  &  150.148527  &  2.017461  &  1.82$^{+0.12}_{0.09}$$^{s}$ &  9.4$^{+0.4}_{0.4}$  &        24.77  &   1.7$^{+0.0}_{0.3}$  & 44.0$^{+1.4}_{1.2}$ &   15.1$\pm$2.5  &     39.70$\pm$0.09  &  43.6  \\
     7787  &  149.678570  &  2.156774  &    0.13$^{+0.00}_{0.00}$$^{s}$ &7.5$^{+0.6}_{0.6}$  &        24.10  &  -1.6$^{+0.4}_{0.6}$  &  42.2$^{+1.0}_{0.9}$ &   22.8$\pm$2.8  &     38.02$\pm$0.01  &  42.4  \\
     7854  &  149.971884  &  2.491272  &  1.66$^{+0.19}_{0.12}$$^{p}$ &  9.1$^{+0.7}_{0.6}$  &        26.31  &   1.6$^{+0.0}_{0.4}$  & 45.0$^{+0.9}_{1.2}$ &   14.0$\pm$2.4  &     39.57$\pm$0.09  &  43.5  \\
     8480  &  150.212280  &  1.691562  &  0.38$^{+0.01}_{0.02}$$^{p}$ &  9.5$^{+0.1}_{0.2}$  &        23.05  &   0.3$^{+0.0}_{0.2}$  & 44.2$^{+0.9}_{0.9}$ &   13.1$\pm$2.4  &      38.0$\pm$0.1  &  42.4  \\
     8692  &  149.696664  &  2.035075  &  0.54$^{+0.02}_{0.02}$$^{p}$ &  9.1$^{+0.4}_{0.4}$  &        24.36  &  -0.1$^{+0.2}_{0.4}$  & 42.6$^{+1.3}_{0.9}$ &   25.6$\pm$2.8  &     38.79$\pm$0.05  &  42.9  \\
     8727  &  149.852717  &  2.493861  &    0.30$^{+0.26}_{0.2}$$^{p}$ &  7.4$^{+0.7}_{0.7}$  &        25.73  &   0.03$^{+0.01}_{0.0}$  & 43.2$^{+0.9}_{0.9}$  & 13.1$\pm$2.4  &     37.8$\pm$0.1  &  42.2  \\
     8788  &  149.439531  &  2.384169  &  0.71$^{+0.06}_{0.06}$$^{p}$ &  8.9$^{+0.6}_{0.6}$  &        25.20  &   0.9$^{+0.1}_{0.3}$  & 44.0$^{+0.9}_{0.9}$ &   49.0$\pm$4.8  &     39.32$\pm$0.05  &  43.3  \\
     9059  &  150.482996  &  2.383103  &  2.63$^{+0.05}_{0.05}$$^{p}$ &  9.3$^{+0.4}_{0.5}$  &        24.76  &   1.6$^{+0.0}_{0.0}$  & 45.2$^{+0.9}_{0.9}$ &   12.5$\pm$2.3  &     40.04$\pm$0.09  &  43.8  \\
     9283  &  150.383343  &  2.449567  &  3.34$^{+0.09}_{0.15}$$^{p}$ &  9.2$^{+0.4}_{0.5}$  &        25.58  &   2.0$^{+0.0}_{0.3}$  & 45.4$^{+0.9}_{0.9}$ &   14.5$\pm$2.8  &     40.3$\pm$0.1  &  44.0  \\
     9814  &  149.698609  &  2.413071  &  1.47$^{+0.35}_{0.25}$$^{p}$ &  8.7$^{+0.8}_{0.6}$  &        26.83  &   1.4$^{+0.1}_{0.1}$  & 		-- 			&   12.1$\pm$2.4  &      39.4$\pm$0.1  &  43.3  \\
     9899  &  150.204635  &  2.212778  &  1.76$^{+0.05}_{0.09}$$^{p}$ &  9.3$^{+0.4}_{0.4}$  &        24.85  &   1.5$^{+0.2}_{0.2}$  & 44.8$^{+1.0}_{1.4}$ &   12.2$\pm$2.4  &      39.6$\pm$0.1  &  43.5  \\
    10054  &  149.981862  &  2.664827  &   1.61$^{+0.12}_{0.1}$$^{p}$ &  9.2$^{+0.5}_{0.5}$  &        25.00  &   1.6$^{+0.0}_{0.3}$  & 45.0$^{+1.0}_{1.2}$ &   11.9$\pm$2.4  &      39.5$\pm$0.1  &  43.4  \\
    10574  &  149.904669  &  2.863645  &   0.34$^{+0.1}_{0.12}$$^{p}$ &  8.1$^{+0.7}_{0.6}$  &        25.89  &  -0.8$^{+0.3}_{0.0}$  & 42.2$^{+1.0}_{0.9}$ &   13.6$\pm$2.7  &     38.07$\pm$0.09  &  42.4  \\
    10629  &  149.483588  &  2.318789  &  0.45$^{+3.48}_{0.19}$$^{p}$ &  7.8$^{+0.5}_{0.5}$  &        26.10  &  -0.6$^{+0.5}_{0.1}$  & 43.2$^{+0.9}_{1.0}$  &   15.5$\pm$3.2  &      38.40$\pm$0.09  &  42.6  \\
    10661  &  150.696557  &  1.985819  &  0.29$^{+0.01}_{0.01}$$^{p}$ &  9.0$^{+0.2}_{0.3}$  &        23.42  &   0.1$^{+0.1}_{0.0}$  &  43.9$^{+0.9}_{0.9}$  &   12.9$\pm$2.7  &      37.8$\pm$0.1  &  42.2  \\ 
    10734  &  149.978025  &  2.848848  &  1.67$^{+0.36}_{0.18}$$^{p}$ &  9.4$^{+0.7}_{0.6}$  &        25.37  &   1.6$^{+0.1}_{0.3}$  & 44.0$^{+1.0}_{0.9}$  &   14.1$\pm$2.9  &      39.6$\pm$0.1  &  43.5  \\
    10896  &  150.069473  &  2.706838  &  2.84$^{+0.18}_{0.17}$$^{p}$ &  9.5$^{+0.5}_{0.5}$  &        25.54  &   1.8$^{+0.2}_{0.1}$  & 45.2$^{+0.9}_{0.9}$  &   12.9$\pm$2.5  &     40.1$\pm$0.1  &  43.8  \\
\hline
\hline
\end{tabular}
\end{minipage}
\raggedright
\smallskip\newline\small {\bf Column designation:}~(1) VLA-COSMOS 3 GHz Large Project ID, (2) right ascension, (3) declination, (4) redshift, $s$ = spectroscopic, $p$ = photometric, (5) stellar mass, (6) i-band magnitude, (7) star formation rate, (8) bolometric luminosity derived from the SED fitting, (9) integrated radio flux at 3 GHz, (10) 1.4 GHz AGN radio luminosity after removing the contribution from star formation, (11) jet power.
\end{table*}

\subsection{Bolometric luminosities}
\begin{figure}
\includegraphics[width=0.48\textwidth]{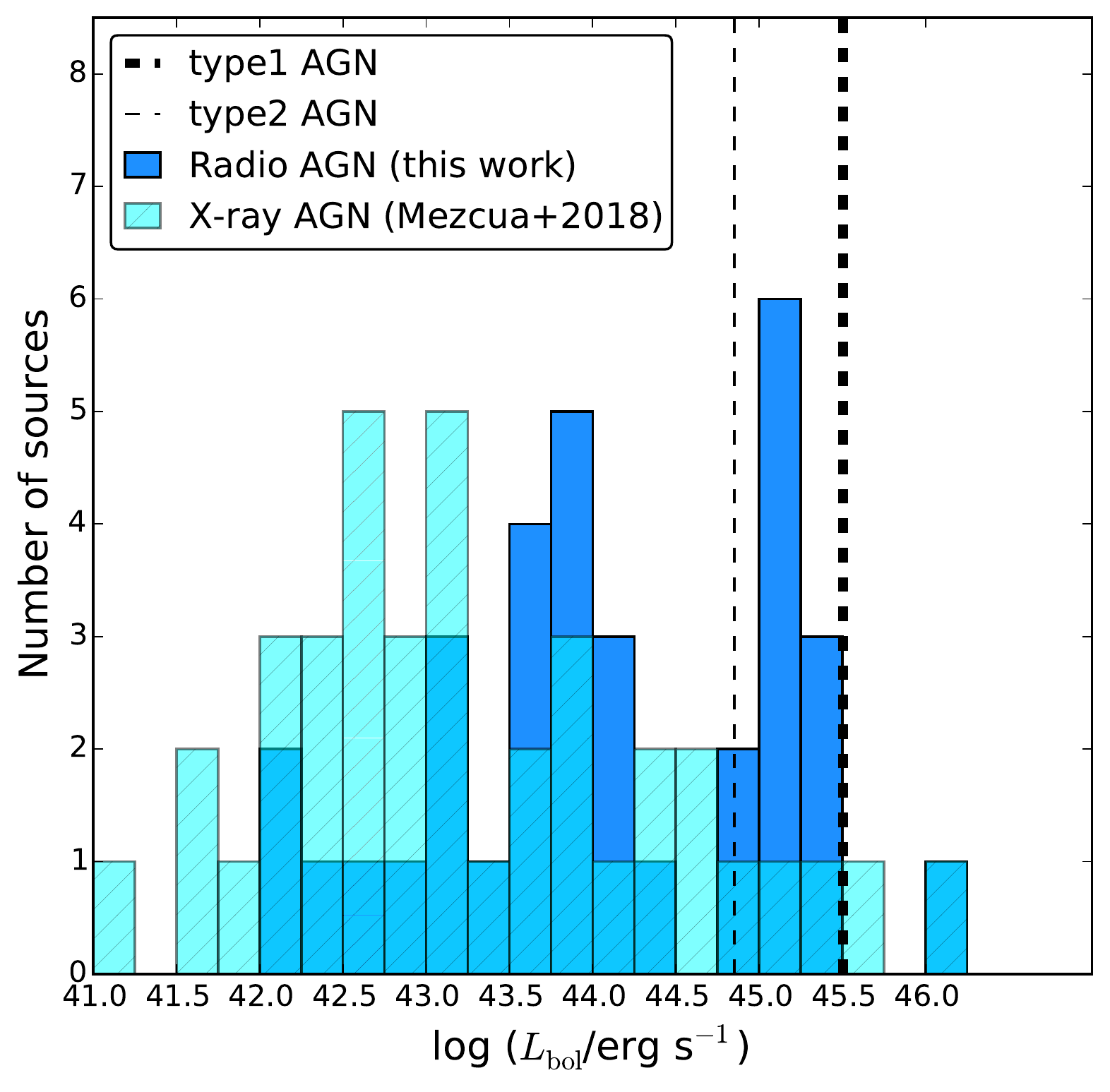}
\includegraphics[width=0.48\textwidth]{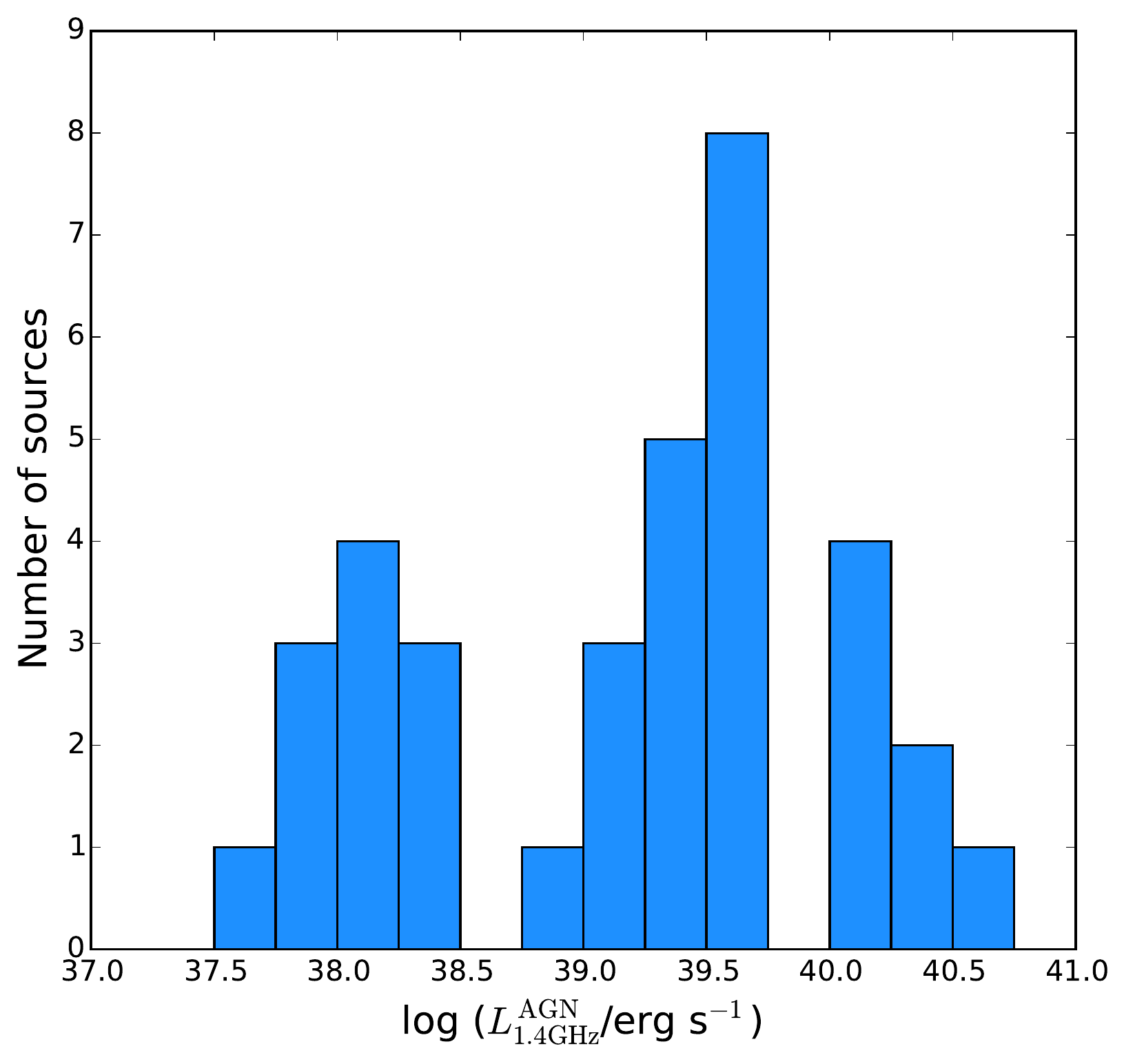}
\protect\caption[figure]{\textbf{Top}. Distribution of bolometric luminosity for the sample of radio AGN dwarf galaxies (blue solid bars) and the X-ray sample of AGN dwarf galaxies from \cite{2018MNRAS.478.2576M} (cyan hashed bars). The vertical dashed lines indicate the median bolometric luminosities of type 1 (thick line) and type 2 (thin line) AGN from \cite{2012MNRAS.425..623L}. \textbf{Bottom}. Distribution of AGN 1.4 GHz radio luminosities for the sample of radio AGN dwarf galaxies.} 
\label{Lbol}
\end{figure}

The bolometric luminosities for the sample of radio AGN are derived from the SED fitting following \cite{2019ApJ...872..168S}. We find values ranging from $L_\mathrm{bol}$ = 1.5 $\times$ 10$^{42}$ erg s$^{-1}$ to 1.2 $\times$ 10$^{46}$ erg s$^{-1}$ and that the distribution presents two peaks (see Fig.~\ref{Lbol}, top). While the lower luminosity peak could be attributed to low-luminosity AGN ($L_\mathrm{bol} \sim 10^{42}$ erg s$^{-1}$; e.g. \citealt{2014ApJ...787...62M}), the main peak is consistent with the typical bolometric luminosity of Seyfert galaxies, quasars, and luminous X-ray selected AGN (e.g. \citealt{2012MNRAS.425..623L}; \citealt{2018arXiv181206025B}), which reinforces the presence of an AGN in the sample of dwarf galaxies. Such double-peaked distribution is also observed in the 1.4 GHz AGN radio luminosities (see Fig.~\ref{Lbol}, bottom), where the lower luminosity of the secondary peak ($L^\mathrm{AGN}_\mathrm{1.4 GHz} \sim 10^{38}$ erg s$^{-1}$) is consistent with that of low-luminosity AGN. The fit of a linear slope between 1.4 GHz AGN radio luminosity and $L_\mathrm{bol}$ yields a slope $m$ = 0.9 $\pm$ 0.1 ($r^2$ = 0.5) and a probability for rejecting the null hypothesis that there is no correlation of $p$ = 1.9 $\times$ 10$^{-6}$. 
The median of the distribution of bolometric luminosities (<$L_\mathrm{bol}$> = 1.0 $\times$ 10$^{44}$ erg s$^{-1}$) is closer to that of type 2 AGN (7.1 $\times$ 10$^{44}$ erg s$^{-1}$; \citealt{2012MNRAS.425..623L}) than type 1 AGN, suggesting that most of the radio AGN might be of type 2.

\subsection{Jet kinetic power and feedback}
The total jet power ($Q_\mathrm{jet}$) of the AGN can be derived from the radio luminosity of the compact core (e.g. \citealt{2007MNRAS.381..589M}; \citealt{2010ApJ...720.1066C}; \citealt{2012MNRAS.423.2498D}). For this we use the correlation between 1.4 GHz radio luminosity and jet kinetic power of \cite{2010ApJ...720.1066C}, most of whose sources are assumed to have a steep spectral index ($\alpha$ = 0.8) as the radio AGN dwarf galaxies here studied (see Sect.~\ref{spectralindex}). A thorough study of the effects of steepening of the spectral index on the $Q_\mathrm{jet}$-$L_\mathrm{1.4GHz}$ relation is performed in \cite{2013ApJ...767...12G}, who find an excellent agreement when comparing samples of FRI and FRII radio galaxies (though see \citealt{2013ApJ...769..129S} for additional effects of source size). Out of the 35 radio AGN dwarf galaxies, 28 show compact radio emission at 3 GHz. For the remaining sources whose 3 GHz radio emission is resolved, the jet power derived should be taken as an upper limit. We obtain jet powers ranging from $Q_\mathrm{jet}$ = 1.2 $\times$ 10$^{42}$ erg s$^{-1}$ to 1.4 $\times$ 10$^{44}$ erg s$^{-1}$, thus up to two orders of magnitude lower than the bolometric luminosities. The values are of the same order as those of AGN in powerful radio galaxies ($Q_\mathrm{jet} \geq 10^{42}$ erg s$^{-1}$; e.g. \citealt{2007MNRAS.381..589M}; \citealt{2014ApJ...787...62M}), which can collimate their radio jets to large (kpc) scales and whose jet mechanical feedback is able to open cavities in the surrounding intergalactic environment (e.g. \citealt{2000MNRAS.318L..65F}; \citealt{2000ApJ...534L.135M}; \citealt{2007ARA&A..45..117M,2012NJPh...14e5023M}; \citealt{2018arXiv181206025B}). Such massive radio galaxies have jet efficiencies, defined as the fraction of accretion energy used in the kinetic motion of the jet ($\eta_\mathrm{jet} = Q_\mathrm{jet}/\eta \dot{M_\mathrm{BH}} c^{2}$, where $\eta$ is the efficiency of conversion of rest mass into energy, $\dot{M}_\mathrm{BH} = L_\mathrm{bol}/\eta c^{2}$ is the accretion rate, and $c$ the speed of light), of $\eta_\mathrm{jet} \geq$ 10\% (e.g. \citealt{2015MNRAS.449..316N}). 

From the $L_\mathrm{bol}$ and $Q_\mathrm{jet}$, we find that 19 of the 35 radio AGN dwarf galaxies have jet efficiencies $\geq$ 10\%, therefore as high as those of massive galaxies whose jet mechanical feedback can inflate cavities in the intracluster medium and prevent and trigger star formation on local (tens of pc) scales in the host galaxy (e.g. \citealt{2014Natur.511..440T}; \citealt{2015A&A...580A...1M}; \citealt{2017Natur.544..202M}). Observational evidence for AGN feedback in dwarf galaxies is frugal (e.g. \citealt{2017ApJ...845...50N}; \citealt{2018MNRAS.476..979P}; \citealt{2019arXiv190509287M}), hence the finding of such a large sample of dwarf galaxies with high jet efficiencies is of significant importance for understanding whether they host the seed BHs of the early Universe. BH accretion rate has been found to positively correlate with SFR in star-forming galaxies out to $z\sim$ 3 (e.g. \citealt{2012ApJ...753L..30M}; \citealt{2013ApJ...773....3C}; \citealt{2015MNRAS.449..373D}; \citealt{2019MNRAS.485.3721Y}). Therefore, if their radio jets trigger star formation, BHs in dwarf galaxies could have significantly grown rather than retained their initial seed BH mass (\citealt{2019NatAs...3....6M}). In this case, the use of the BH occupation fraction in dwarf galaxies as a tool to distinguish between seed BH formation models should be revisited (e.g. \citealt{2008MNRAS.383.1079V}; \citealt{2010MNRAS.408.1139V}; \citealt{2018MNRAS.478.2576M}).

\subsection{Black hole mass and accretion rate}
AGN in dwarf galaxies are typically found to have near- to super-Eddington accretion rates (e.g. \citealt{2007ApJ...670...92G}; \citealt{2016ApJ...829...57B}; \citealt{2018MNRAS.478.2576M}), even those exhibiting radio outflows (e.g. \citealt{2006ApJ...646L..95W}; \citealt{2008ApJ...686..892T}; \citealt{2008ApJ...686..838W}; \citealt{2018MNRAS.478.2576M}). Assuming an Eddington ratio $\lambda_\mathrm{Edd} =$ 1 and using $M_\mathrm{BH}$ = $L_\mathrm{bol}$/($\lambda_\mathrm{Edd} \times 1.3 \times 10^{38}$), the radio AGN dwarf galaxies would have BH masses ranging from log ($M_\mathrm{BH}$/M$_{\odot}$) = 4.1 to 7.9 (see Table~\ref{table2}). 

These BH masses are slightly higher than those that would be obtained (log ($M_\mathrm{BH}$/M$_{\odot}$) = 3.6 to 5.9) applying the $M_\mathrm{BH}-M_\mathrm{*}$ correlation from \cite{2015ApJ...813...82R} drawn from local AGN and which includes a sample of dwarf galaxies with broad-line AGN, suggesting that most of the radio AGN dwarf galaxies might host IMBHs. However, the BH-galaxy scaling relations are found to have a break and a large scatter at the low-mass end (e.g. \citealt{2011ApJ...742...68J}; \citealt{2017IJMPD..2630021M}; \citealt{2018ApJ...855L..20M}; \citealt{2015ApJ...798...54G}; \citealt{2018arXiv181004888D}; \citealt{2019MNRAS.485.1278S}), so they are not reliably applicable to galaxies with low stellar mass. A flattening of the low-mass end (e.g. \citealt{2011ApJ...742...68J}; \citealt{2017IJMPD..2630021M}; \citealt{2018ApJ...855L..20M}; \citealt{2019MNRAS.485.1278S}) would be consistent with the predictions from models of direct collapse formation of seed BHs (e.g. \citealt{2009MNRAS.400.1911V}; \citealt{2010MNRAS.408.1139V}; but see \citealt{2018MNRAS.tmp.2337R}) or could be ascribed to a bimodality in the accretion efficiency of BHs in which low-mass BH accrete inefficiently (\citealt{2018ApJ...864L...6P}), while a steepening at low stellar masses (e.g. \citealt{2015ApJ...798...54G}; \citealt{2018arXiv181004888D}) would imply that the radio AGN dwarf galaxies have all BH masses $\leq 10^{3}$ M$_{\odot}$.

In the case of AGN with jet radio emission the BH mass can be estimated using the fundamental plane of BH accretion, which is an empirical correlation, reinforced by theoretical accretion models, between BH mass, nuclear X-ray luminosity and core radio luminosity valid from stellar-mass to supermassive BHs (e.g. \citealt{2003MNRAS.345.1057M}; \citealt{2004A&A...414..895F}; \citealt{2006A&A...456..439K}; \citealt{2009ApJ...706..404G,2014ApJ...788L..22G}; \citealt{2012MNRAS.419..267P}; \citealt{2018A&A...616A.152S}; \citealt{2019ApJ...871...80G}). To estimate the BH masses for the radio AGN dwarf galaxies we use the fundamental plane of \cite{2009ApJ...706..404G}, which has a scatter of 0.77 dex and has been proven in the IMBH regime:
\begin{equation}
\label{gultekin}
\begin{split}
\mathrm{log} L_\mathrm{R} = (4.80 \pm 0.24) + (0.78 \pm 0.27) \mathrm{log} M_\mathrm{BH} + \\ (0.67 \pm 0.12) \mathrm{log} L_\mathrm{X}
\end{split}
\end{equation}
where $L_\mathrm{R}$ is the 5 GHz radio luminosity and $L_\mathrm{X}$ the 2-10 keV X-ray luminosity. None of the radio AGN dwarf galaxies are detected in X-rays at a significant level in the \textit{Chandra} COSMOS-Legacy survey; hence to compute the X-ray luminosities we make use of the \textit{Chandra} stacking analysis tool CSTACK\footnote{\url{http://lambic.astrosen.unam.mx/cstack/}} v4.32. CSTACK stacks the \textit{Chandra} data at the position of each radio AGN dwarf galaxy, removing any nearby sources which are X-ray detected or resolved, and returns the source counts, background counts, exposure time, and count rate, among other variables, for each stacked field as well as for the population of input galaxies in the soft (0.5-2 keV) and hard (2-10 keV) bands. A detailed description of the method can be found in \cite{2016ApJ...817...20M}. For the purposes of this paper we take the source counts in the stacked field of each radio AGN dwarf galaxy in the 2-10 keV band. These are available for 34 of the 35 radio AGN dwarf galaxies. The remaining source is out of the \textit{Chandra} COSMOS-Legacy survey coverage. Because of the low number of counts, we use Gehrels statistics (\citealt{1986ApJ...303..336G}) to compute upper values of the count rate for each dwarf galaxy. Following the \textit{Chandra} COSMOS-Legacy survey, the count rates are converted to fluxes assuming a photon index $\Gamma$ = 1.4 and a Galactic column density $N_\mathrm{H}$ = 2.6 $\times$ 10$^{20}$ cm$^{-2}$ (\citealt{2016ApJ...819...62C}). The corresponding upper limits on the K-corrected 2-10 keV X-ray luminosities for the 34 radio AGN dwarf galaxies with stacked X-ray emission range from $L_\mathrm{2-10\ keV}$ = 6.4 $\times$ 10$^{40}$ to 8.4 $\times$ 10$^{43}$ erg s$^{-1}$ (see Table~\ref{table2}). From these values and the 5 GHz AGN radio luminosities obtained in the previous section we find that the BH masses derived from the fundamental plane of BH accretion (eq.~\ref{gultekin}) range from log ($M_\mathrm{BH}$/M$_{\odot}$) = 5.6 to 7.9. We note that this range of BH masses is consistent with that derived assuming $\lambda_\mathrm{Edd} =$ 1 despite the fundamental plane should only be applied for sources accreting at sub-Eddington rates (e.g. \citealt{2012MNRAS.419..267P}; \citealt{2018MNRAS.474.1342M}).

\begin{table}
\centering
\caption{Radio AGN dwarf galaxies: ID, BH mass (derived from $L_\mathrm{bol}$ assuming $\lambda_\mathrm{Edd}$ = 1), stacked X-ray luminosity, and ratio of radio to X-ray emission: $R_\mathrm{X}$ = $\nu L_{\nu}$(5 GHz)/$L_\mathrm{X}$(2-10 keV). $\dagger$ No $L_\mathrm{6\mu m}$ is available.}
\label{table2}
\begin{tabular}{lcccc}
\hline
\hline 
  ID$\_$VLA &        log $M_\mathrm{BH}$	&   log $L_\mathrm{X}$ &  log $R_\mathrm{X}$  \\
  	 	 &          (M$_\mathrm{\odot}$) 		&   (erg s$^{-1}$)	  	&      				\\
 \hline
      515  &            5.5  &          $<$42.1  &  $>$-3.0  \\
      909  &            5.7  &          $<$41.9  &  $>$-3.0  \\
     1716  &            6.0  &          $<$42.9  &  $>$-3.2  \\
     1809  &            5.4  &          $<$41.7  &  $>$-3.7  \\
     2329  &            5.6  &          $<$42.5  &  $>$-3.4  \\
     2619  &            8.0  &          $<$43.0  &  $>$-4.1  \\
     2867  &            4.2  &          $<$41.2  &  $>$-3.4  \\
     2945  &            6.1  &          $<$42.9  &  $>$-4.3  \\
     3295  &            6.4  &          $<$43.1  &  $>$-3.8  \\
     3317  &            4.7  &          $<$41.4  &  $>$-3.9  \\
     3770  &            5.0  &          $<$42.7  &  $>$-3.9  \\
     3896  &            7.4  &          $<$43.9  &  $>$-3.8  \\
     4646  &            6.9  &          $<$43.6  &  $>$-3.9  \\
     5871  &            7.2  &          $<$43.5  &  $>$-4.3  \\
     5931  &            7.0  &          $<$43.5  &  $>$-3.8  \\
     5973  &            5.6  &          $<$41.2  &  $>$-4.0  \\
     6775  &            5.2  &          $<$42.1  &  $>$-4.1  \\
     7123  &            5.9  &          $<$43.2  &  $>$-3.9  \\
     7787  &            4.1  &          $<$40.8  &  $>$-3.2  \\
     7854  &            6.9  &          $<$43.3  &  $>$-4.1  \\
     8480  &            6.1  &          $<$41.9  &  $>$-5.5  \\
     8692  &            4.5  &          $<$42.2  &  $>$-3.8  \\
     8727  &            5.1  &          $<$41.9  &  $>$-4.4  \\
     8788  &            5.9  &          $<$42.7  &  $>$-3.8  \\
     9059  &            7.1  &          $<$43.7  &  $>$-4.1  \\
     9283  &            7.3  &          $<$43.9  &  $>$-4.0  \\
     9814  &            - $\dagger$  &          $<$43.2  &  $>$-4.2  \\
     9899  &            6.7  &          $<$43.3  &  $>$-4.1  \\
    10054  &            6.9  &          $<$43.1  &  $>$-4.1  \\
    10574  &            4.1  &          $<$42.0  &  $>$-4.3  \\
    10629  &            5.0  &          $<$42.6  &  $>$-4.6  \\
    10661  &            5.8  &          $<$41.5  &  $>$-4.1  \\
    10734  &            5.9  &          $<$43.4  &  $>$-4.2  \\
    10896  &            7.1  &          $<$43.9  &  $>$-4.2  \\
\hline
\hline
\end{tabular}
\end{table}

A more robust fundamental plane than that of \cite{2009ApJ...706..404G} has been recently published by \cite{2019ApJ...871...80G}. \cite{2019ApJ...871...80G} provide a mass estimator that should be most useful for distinguishing between XRBs, IMBHs and supermassive BHs and that should be applied only in the case of having both a radio and an X-ray detection:
\begin{equation}
\label{gultekin2019}
\mathrm{log} M_\mathrm{BH} = (0.55 \pm 0.22) + (1.09 \pm 0.10) \mathrm{log} L_\mathrm{R} + \\ (-0.59^{+0.16}_{-0.15}) \mathrm{log} L_\mathrm{X}
\end{equation}
Although the radio AGN dwarf galaxies studied here are not X-ray detected, we apply the correlation from \cite{2019ApJ...871...80G} using the upper limits on the X-ray luminosities derived from the stacking. We find that the (upper limits) on the BH mass for the 34 radio AGN dwarf galaxies with stacked X-ray emission range from log ($M_\mathrm{BH}$/M$_{\odot}$) = 5.6 to 8.8, which are consistent within the scatter with the range of masses derived using the correlation of \cite{2009ApJ...706..404G}.

Assuming that the radio AGN accrete at sub-Eddington rates (e.g. $\lambda_\mathrm{Edd}$ = 10$^{-3}$) 43\% of the sources would have BH masses of 10$^{9}$-10$^{10}$ M$_{\odot}$, which are unreasonably large for dwarf galaxies but more typical of supermassive BHs in brightest cluster galaxies (e.g. \citealt{2011Natur.480..215M}; \citealt{2018MNRAS.474.1342M}). We thus conclude that the radio AGN dwarf galaxies host a mixture of mostly IMBHs, but possibly also some supermassive BHs with log ($M_\mathrm{BH}$/M$_{\odot}$) $<$ 8, accreting at near- to super-Eddington rates. This would be consistent with formation models of direct collapse seed BHs, which are predicted to accrete at super-Eddington rates and to emit violent outflows in the form of bipolar jets (\citealt{2019MNRAS.486.3892R}). The presence of radio jets in systems accreting close to or above Eddington rates has been also observed in microquasars such as GRS 1915+105 and in ultraluminous X-ray sources (e.g. \citealt{2004ARA&A..42..317F}; \citealt{2013Natur.493..187M}; \citealt{2014Sci...343.1330S}; \citealt{2014MNRAS.439L...1C,2015MNRAS.452...24C}; \citealt{2019arXiv190601597V}). 

A light seed BH of $\sim10^2$ M$_{\odot}$ accreting at super-Eddington rates could increase its BH mass by one order of magnitude in just 0.5 Myr (e.g. Fig. 3 in \citealt{2016MNRAS.456.2993L}; Fig. 1 in \citealt{2017A&G....58c3.22S}). Such process could occur in dwarf galaxies hosting the relics of the early Universe seed BHs: given that AGN switch on and off in phases that can last $\sim10^5$ yr (e.g. \citealt{2015MNRAS.451.2517S}), the finding of near- to super-Eddington accreting low-mass AGN in dwarf galaxies suggests that their BHs could be the leftovers of light seeds that have grown into heavy seed BHs (e.g. \citealt{2019NatAs...3....6M}). This growth could be accelerated by mergers and feedback processes and it could explain why no 'light' BHs of less than $10^4-10^5$ M$_{\odot}$ are found in dwarf galaxies despite light seeds being more abundant in the early Universe than heavy seeds (\citealt{2019NatAs...3....6M}).

From the 5 GHz radio luminosity and upper limits on the 2-10 keV X-ray luminosity we can also derive the ratio of radio to X-ray emission: $R_\mathrm{X}$ = $\nu L_{\nu}$(5 GHz)/$L_\mathrm{X}$(2-10 keV) (\citealt{2003ApJ...583..145T}). XRBs have typically log $R_\mathrm{X} <$ -5.3, IMBHs -5.3 $<$ log $R_\mathrm{X} <$ -3.8, low-luminosity AGN -3.8 $<$ log $R_\mathrm{X} <$ -2.8, radio-loud type 1 quasars log $R_\mathrm{X} >$ -3, and SNRs log $R_\mathrm{X} \sim$ -2 (e.g. \citealt{2013MNRAS.436.1546M,2013MNRAS.436.2454M}; \citealt{2003ApJ...583..145T}; \citealt{2014arXiv1408.1090H}). The $R_\mathrm{X}$ ratios for the 34 radio AGN with stacked X-ray emission range from log $R_\mathrm{X}$ = -5.5 to -3.0 and are thus consistent with those expected from IMBHs (see Table~\ref{table2}). Given the upper limits on the X-ray luminosity, the values of $R_\mathrm{X}$ should be considered as lower limits, which rules out an XRB nature and reinforces the AGN origin of the radio emission. 


For low-mass AGN, $L_\mathrm{X}/L_\mathrm{bol}$ is typically $\sim$0.1 (e.g. \citealt{2004MNRAS.351..169M}; the mean $L_\mathrm{X}/L_\mathrm{bol}$ is 0.09 in the sample of X-ray detected dwarf galaxies of \citealt{2018MNRAS.478.2576M}). Since for the radio AGN dwarf galaxies studied here $L_\mathrm{bol} \sim 10^{42}-10^{46}$ erg s$^{-1}$, taking $L_\mathrm{X}/L_\mathrm{bol}$ = 0.1 we would expect X-ray luminosities of $\sim10^{41}-10^{45}$ erg s$^{-1}$. Instead, the upper limits derived from the X-ray stacking are one to two orders of magnitude lower for four of the radio AGN dwarf galaxies. Taking the BH mass derived from the fundamental plane, assuming an Eddington ratio $\sim$1 and that $L_\mathrm{X}$ = 0.1 $\times L_\mathrm{bol}$ we would also expect X-ray luminosities one to two orders of magnitude higher than the observed upper limits. This suggests that these four radio AGN dwarf galaxies could be heavily obscured and their $R_\mathrm{X}$ ratios consistent with low-luminosity AGN rather than IMBHs.

\section{Conclusions and open issues}
\label{conclusions}
Because of their low-mass and presumably quiet accretion and merger history, dwarf galaxies are in the spotlight of studies seeking for the leftover of the early Universe seed BHs. Whether these seeds have remained untouched or have not significantly grown since their formation is however unclear. Dwarf galaxies are found to undergo numerous mergers (e.g. \citealt{2010MNRAS.406.2267F}) and to possibly suffer the impact of AGN/BH feedback (e.g. \citealt{2016MNRAS.463.2986S}; \citealt{2018MNRAS.473.5698D}; \citealt{2018arXiv180704768B}; \citealt{2018MNRAS.476..979P}; \citealt{2019arXiv190509287M}), both of which processes would affect the growth of the seed BH (\citealt{2014ApJ...794..115D}; \citealt{2019MNRAS.486.3892R}; \citealt{2019NatAs...3....6M}).

To investigate the presence of AGN feedback in dwarf galaxies, in this paper we have searched for radio AGN in a sample of dwarf galaxies drawn from the VLA-COSMOS 3 GHz Large Project. After accounting for the contribution from star formation to the radio emission, we find 35 radio AGN dwarf galaxies with AGN radio luminosities in the range $L^\mathrm{AGN}_\mathrm{1.4 GHz} \sim 10^{37}-10^{40}$ erg s$^{-1}$. Among this sample there are four sources at $z >$ 2.5, including the highest-redshift dwarf galaxy found to have an AGN (ID\_VLA 3896, $z$ = 3.4), which proves the power of deep radio surveys in detecting AGN at high redshifts. 

The radio AGN dwarf galaxies have bolometric luminosities $\gtrsim 10^{42}$ erg s$^{-1}$, which reinforces their AGN nature. The radio emission of most of the sources is compact at 3 GHz, indicating that its origin is most likely from a radio jet. The jet powers derived from the AGN radio luminosities range from $Q_\mathrm{jet} \sim 10^{42}$ to 10$^{44}$ erg s$^{-1}$ and are thus of the same order as those of powerful radio galaxies (e.g. \citealt{2007MNRAS.381..589M}; \citealt{2014ApJ...787...62M}). The jet efficiencies for more than 50\% of the sources are also found to be as high ($\geq 10$\%) as those of massive galaxies. In massive radio galaxies, jet mechanical feedback is found to inflate cavities in the intergalactic environment and to both hamper and trigger star formation in the host galaxy (e.g. \citealt{2007ARA&A..45..117M,2012NJPh...14e5023M}; \citealt{2014Natur.511..440T}; \citealt{2015A&A...580A...1M}; \citealt{2017Natur.544..202M}). The finding that dwarf galaxies host radio AGN with jets as powerful as those of massive galaxies indicates that AGN feedback can also have a strong impact in these low-mass galaxies, and thus possibly affect the material available for the BH to grow. 

Several IMBHs with powerful jet radio emission have been found as off-nuclear sources, possibly the remnant core of a dwarf galaxy that was stripped in the course of a minor merger (e.g. \citealt{2012Sci...337..554W}; \citealt{2013MNRAS.436.3128M,2015MNRAS.448.1893M,2018MNRAS.480L..74M}); however, observational evidence for AGN feedback being significant in dwarf galaxies was so far scarce (e.g. \citealt{2017ApJ...845...50N}; \citealt{2018ApJ...861...50B}; \citealt{2018MNRAS.476..979P}; \citealt{2019arXiv190201401D}; \citealt{2019arXiv190509287M}). The discovery of such a large sample of dwarf galaxies with radio AGN as energetic as those of massive galaxies has thus important implications for models of seed BH formation: if AGN radio jets in dwarf galaxies have a significant impact on their hosts, e.g. triggering star formation, and possibly on the growth of their BHs, then the low-mass AGN found in dwarf galaxies should not be treated as relics of the early Universe seed BHs. 

High-resolution kinematical studies would be needed in order to probe the effects of the radio jet on the surrounding environment of its host dwarf galaxy, which is currently only feasible for the closest and brightest dwarf galaxies. The next generation of major facilities such as the \textit{James Webb Space Telescope} will provide a leap forward in this respect, allowing us to probe in detail the impact of the radio jets in dwarf galaxies on their hosts.

\section*{Acknowledgments}
The authors thank the anonymous referee for insightful comments. 
M.M. acknowledges support from the Spanish Juan de la Cierva program (IJCI-2015-23944). The authors thank Alister Graham, John Regan, Francesco Shankar, and Marta Volonteri for insightful discussions.


\bibliographystyle{mnras}
\bibliography{/Users/mmezcua/Documents/referencesALL}


\bsp	
\label{lastpage}
\end{document}